\pdfoutput=1
\documentclass[a4paper]{article}
\usepackage[a4paper,margin=3.08cm]{geometry}
\usepackage{amssymb,amsmath}
\usepackage{tablefootnote}
\usepackage{upgreek}
\usepackage{graphicx}
\usepackage{placeins}
\usepackage{algorithmic}
\algsetup{indent=1.5em}
\usepackage{tikz}
\usetikzlibrary{shapes.geometric}
\usetikzlibrary{trees}
\usetikzlibrary{arrows}
\newcommand\tstrut{\rule{0pt}{2.4ex}}
\newcommand\bstrut{\rule[-1.0ex]{0pt}{0pt}}

\usepackage{hyperref}
\hypersetup{
    pdfauthor={Nathan Clisby},
    pdftitle={Scale-free Monte Carlo method for calculating the critical exponent gamma of self-avoiding walks},
    breaklinks = true, colorlinks = true,
    linkcolor = blue,
    anchorcolor = blue,
    citecolor = blue,
    filecolor = blue,
    urlcolor = blue
    }

\providecommand{\url}[1]{{#1}}

\expandafter\ifx\csname urlstyle\endcsname\relax
  \else
  \fi

\newcommand{\Z}{{\mathbb Z}}

\newcommand{\tint}{\tau_{\mathrm{int}}}

\newcommand{\DDE}{D_{\mathrm{E}}}
\newcommand{\DDG}{D_{\mathrm{G}}}

\def\root{\draw[fill] +(0,0) circle (5pt);}
\def\sww{\draw ++(0,0)}

\def\sdw{\draw[style=densely dotted] ++(0,0)}
\def\ew{;}
\def\r{-- ++(1,0)}
\def\l{-- ++(-1,0)}
\def\u{-- ++(0,1)}
\def\d{-- ++(0,-1)}

\newcommand{\btn}{\ensuremath{\widetilde{B}_N}}
\newcommand{\bn}{\ensuremath{B_N}}

\begin{document}
\title{Scale-free Monte Carlo method for calculating the critical
exponent $\gamma$ of self-avoiding walks
}

\author{Nathan Clisby \\
School of Mathematics and Statistics, \\
The University of Melbourne, Victoria 3010, Australia. \\
nclisby@unimelb.edu.au}

\date{January 29, 2017}

\maketitle

\begin{abstract}
We implement a scale-free version of the pivot algorithm and use it to sample pairs of three-dimensional self-avoiding walks, for the purpose of efficiently calculating an observable that corresponds to the probability that pairs of self-avoiding walks remain self-avoiding when they are concatenated. We study the properties of this Markov chain, and then use it to find the critical exponent $\gamma$ for self-avoiding walks to unprecedented accuracy. Our final estimate for $\gamma$ is $1.15695300(95)$. 

\bigskip
\noindent \textbf{Keywords} self-avoiding walk; critical exponent;
Monte Carlo; pivot algorithm 
\end{abstract}

\section{Introduction}
\label{sec-introduction}

An $N$-step self-avoiding walk (SAW) on the $d$-dimensional cubic lattice is a mapping $\omega : \{0,1,\ldots, N\} \to
\Z^d$ with $|\omega(i+1)-\omega(i)|=1$ for each $i$ ($|x|$ denotes the
Euclidean norm of $x$), with $\omega(0)$ at the origin, and with
$\omega(i) \neq \omega(j)$ for all $i \neq j$.
It is of fundamental interest in the theory of critical phenomena as the
$n \rightarrow 0$ limit of the $n$-vector model, and
is the simplest model which captures the universal behavior of polymers
in a good solvent.

The number of self-avoiding walks of
length $N$ on $\Z^3$, which we denote $c_N$, is believed to be given by 
\begin{align} 
    c_N &= A N^{\gamma - 1} \mu^N \left(1 + \frac{a}{N^{\Delta_1}} +
O\left(\frac{1}{N}\right) \right).
\label{eq:asymptotic}
\end{align}
The exponents $\gamma$ and $\Delta_1$ are universal, i.e.
they are dependent only on the dimensionality of the lattice,
while the growth constant $\mu$ and amplitude $A$ are not.
The exponent $\Delta_1 =
0.528(8)$~\cite{Clisby2016HydrodynamicRadiusForSAWs}, and next-to-leading correction
terms with exponents $-1, -2\Delta_1, -\Delta_2$ are folded into the
$O(1/N)$ expression. For bipartite lattices there is an additional
``anti-ferromagnetic'' term which has a factor of $(-1)^N$. It is important
to take this into account when studying series from exact
enumeration~\cite{Clisby2007Selfavoidingwalk}, but it is negligible for the values of $N$
that are accessible to the Monte Carlo computer experiments considered
here and so we neglect it.

In two dimensions the critical exponent $\gamma$ is known exactly,
predicted to be $43/32$ over thirty years ago via Coulomb gas arguments
by Nienhuis~\cite{Nienhuis1982ExactCriticalPoint}. This has been
verified to extremely high precision via enumerations using the finite
lattice
method~\cite{Neef1977Seriesexpansionsfrom,Conway1993Algebraictechniquesenumerating,Conway1996SquareLatticeSelf,Jensen2004Enumerationselfavoiding,Jensen2013newtransfermatrixArxiv};
the most recent estimate confirms the exact result to more than five
decimal places, $\gamma =
1.343745(5)$~\cite{Jensen2013newtransfermatrixArxiv}.

In three dimensions the finite lattice method is not as powerful, and
the best estimates for $\gamma$ come from other enumeration
techniques~\cite{Schram2011ExactEnumerationsSelfAvoidingWalksThesis} and Monte Carlo
simulation~\cite{Caracciolo1992JoinandCut,Caracciolo1998Highprecisiondetermination}.

In this work, we will calculate the critical exponent $\gamma$ and
amplitude $A$ for SAWs on $\Z^3$ to a high degree of accuracy via a
Monte Carlo computer experiment.  We will use an efficient
implementation of the pivot
algorithm~\cite{Clisby2010Efficientimplementationpivot,Clisby2010AccurateEstimateCritical}
which makes it feasible to rapidly sample self-avoiding walks of
millions of steps.  Our simulation framework is similar to an earlier
calculation of the growth constant
$\mu$~\cite{Clisby2013CalculationConnectiveConstant}; here we go into
more depth and explicitly study the behaviour of the autocorrelation
function of the Markov chain.

\subsection{Outline of paper}
\label{subsec:outline}

We introduce our method to calculate $\gamma$ in Sec.~\ref{sec:method},
which includes a calculation of the autocorrelation function of the Markov chain for different
choices of sampling scheme. We then 
present our results and analysis in Sec.~\ref{sec:results}. Finally, we
compare our estimates for $\gamma$ and $A$ to values from the
literature,
discuss the potential for scale-free moves as a paradigm for modeling
polymers, and give a brief conclusion in Sec.~\ref{sec:discussion}.

\section{Method}
\label{sec:method}

\subsection{An observable for calculating $\gamma$ via the pivot
algorithm}

The pivot algorithm is the most efficient method known for sampling
self-avoiding
walks~\cite{Lal1969MonteCarlocomputer,Madras1988PivotAlgorithmHighly},
and recent
improvements~\cite{Kennedy2002fasterimplementationpivot,Clisby2010AccurateEstimateCritical,Clisby2010Efficientimplementationpivot}
have made it even more effective, especially in the large $N$ limit.
These improvements are highly beneficial as they allow one to obtain
accurate data for large $N$, which reduces systematic errors due to
corrections-to-scaling.

The method described here is very similar to that of a recent
paper~\cite{Clisby2013CalculationConnectiveConstant}, but as we wish to
emphasize different aspects of the method we will keep the description
self-contained, even though this will result in a degree of repetition.

The principal difficulty in applying the pivot algorithm to the
estimation of $\gamma$ is that it samples walks in the fixed-length
ensemble, whereas $\gamma$ is intrinsically associated with the growth
in the number of walks as a function of length.  Caracciolo et
al.~\cite{Caracciolo1992JoinandCut} overcame this difficulty by
inventing the join-and-cut algorithm which samples pairs of
self-avoiding walks of fixed total length. Another approach is the
Berretti-Sokal algorithm~\cite{Berretti1985NewMonteCarlo} which
naturally samples walks of different lengths.

We wish to use the pivot algorithm to sample self-avoiding walks, and so
we must find an observable that allows us to estimate $\gamma$ from
the fixed-length ensemble. 

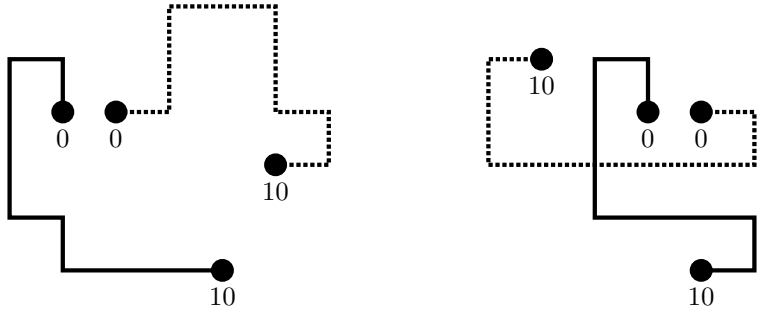
\begin{figure}[htb]
\begin{center}
\begin{tikzpicture}[ultra thick,scale=0.70]
\root
\sww \u \l \d \d \d \r \d \r \r \r \ew
\node at (0,-0.5) {0};
\node at (3.0,-3.5) {10};
\draw[fill] +(3,-3) circle (5pt);
\begin{scope}[shift={(1,0)}]
\root
\sdw \r \u \u \r \r \d \d \r \d \l \ew
\node at (0,-0.5) {0};
\node at (3,-1.5) {10};
\draw[fill] +(3,-1) circle (5pt);
\end{scope}
\begin{scope}[shift={(11,0)}]
\root
\sww \u \l \d \d \d \r \r \r \d \l \ew
\node at (0,-0.5) {0};
\node at (1,-3.5) {10};
\draw[fill] +(1,-3) circle (5pt);
\begin{scope}[shift={(1,0)}]
\root
\sdw \r \d \l \l \l \l \l \u \u \r \ew
\node at (0,-0.5) {0};
\node at (-3,0.5) {10};
\draw[fill] +(-3,1) circle (5pt);
\end{scope}
\end{scope}
\end{tikzpicture}
\end{center}
\caption{Examples of the concatenation of two walks of ten steps on the square
lattice, with the labels of the innermost and outermost sites shown. For the pair of walks on the left $B(\omega_1,\omega_2) = 1$,
while on the right $B(\omega_1,\omega_2) = 0$.\label{fig:concatenation}}
\end{figure}

To do this we sample the same observable as a previous
paper~\cite{Clisby2013CalculationConnectiveConstant}: the probability
that two self-avoiding walks of length $N$ can be concatenated to form a
self-avoiding walk of length $2N+1$.  We note that the use of pairs of
walks to estimate $\gamma$ was suggested by Madras and
Sokal~\cite{Madras1988PivotAlgorithmHighly}, and that the join-and-cut
algorithm~\cite{Caracciolo1992JoinandCut} is also similar.  We define an
observable $B$ on pairs of walks
$\omega_1$ and $\omega_2$ via
\begin{align}
    B(\omega_1,\omega_2) &= \begin{cases} 0 & \text{if $\omega_1 \circ
\omega_2$ not self-avoiding} \\
 1 & \text{if $\omega_1 \circ \omega_2$ self-avoiding} \end{cases} 
 \end{align}
The concatenation operation is illustrated in
Fig.~\ref{fig:concatenation}; it is not the standard operation because an additional
bond is inserted between $\omega_1$ and $\omega_2$.
We adopt the convention that the sites of the walk which are incident to
the concatenating bond
are labeled 0, and increase in number going out to the free ends.

We define $\bn$ as the expectation of $B$ on the set of all pairs of
self-avoiding walks of $N$
steps:
\begin{align}
    \bn &\equiv \langle B(\omega_1,\omega_2) \rangle_{|\omega_1|=N, |\omega_2|=N} 
\\ &= \frac{1}{c_N c_N} \sum_{|\omega_1|=N, |\omega_2|=N} B(\omega_1,\omega_2) \\
&= \frac{c_{2N+1}}{\Omega c_N^2},
\end{align}
where $\Omega$ is the coordination number of the lattice ($\Omega =
6$ for the simple cubic lattice).
Now we define $\btn \equiv \Omega \bn$, and use the
asymptotic form of $c_N$ from (\ref{eq:asymptotic}) to obtain:
\begin{align}
    \btn  
&= \frac{c_{2N+1}}{c_N^2}
    \\ &=  \frac{A (2N+1)^{\gamma-1} \mu^{2N+1} \left(1 +
    \frac{a}{(2N+1)^{\Delta_1}} + O\left(\frac{1}{N}\right)
\right)}{A^2 N^{2\gamma-2}\mu^{2N} \left(1
+ \frac{a}{N^{\Delta_1}} +
O\left(\frac{1}{N}\right)\right)^2}\\
&= \frac{2^{\gamma-1}\mu}{A} N^{1-\gamma} \left(1 +
\frac{b}{N^{\Delta_1}} + O\left(\frac{1}{N}\right)\right).
\label{eq:bn} 
\end{align}

\subsection{The pivot algorithm for sampling self-avoiding walks}

The pivot algorithm is a Markov chain Monte Carlo algorithm which
samples walks of fixed length $N$.  The elementary move is a
\emph{pivot}, where a lattice symmetry operation (rotation or
reflection) is applied to part of a walk, and it generates a correlated
sequence of self-avoiding walks as follows:
\begin{enumerate}
\item Select a pivot site of the current SAW according to some
    prescription (usually uniformly at random, here we will use a
        non-uniform distribution);
\item Randomly choose a lattice symmetry (rotation or reflection) which is not the identity;
\item Apply this symmetry to one of the two sub-walks created by splitting
the walk at the pivot site;
\item If the resulting walk is self-avoiding: {\em accept} the pivot
and update the configuration;
\item If the resulting walk is not self-avoiding: {\em reject} the pivot
and keep the old configuration;
\item Repeat.
\end{enumerate}
The pivot algorithm is ergodic, and satisfies the detailed balance
condition which ensures that self-avoiding walks are sampled uniformly at
random~\cite{Madras1988PivotAlgorithmHighly}.

Successful pivot moves make large changes to global observables which
measure the size of a walk, and Madras and
Sokal~\cite{Madras1988PivotAlgorithmHighly} argued that in fact the
integrated autocorrelation time, $\tint$, for such observables was of
the same order as the mean time for a successful pivot to occur.  For
the simple cubic lattice the probability of a pivot move being
successful is $O(N^p)$ with $p \approx 0.11$, which leads to $\tint =
O(N^p)$ for global observables.

Madras and Sokal~\cite{Madras1988PivotAlgorithmHighly} gave a hash table
implementation of the pivot algorithm which resulted in mean CPU time of
$O(N^{1-p})=O(N^{0.89})$ per pivot attempt (alternatively, CPU time
$O(N)$ per successful pivot). This has since been improved by
Kennedy~\cite{Kennedy2002fasterimplementationpivot} to roughly mean CPU
time of $O(N^{0.74})$ per pivot attempt, and further still by the
present
author~\cite{Clisby2010Efficientimplementationpivot,Clisby2010Efficientimplementationpivot}
to $O(\log N)$. This makes the pivot algorithm extremely efficient for
sampling global observables for self-avoiding walks, but it is not
obvious how efficient it is for sampling our observable $B$.

\subsection{Autocorrelation functions for uniform and scale-free pivot moves}

We now proceed to calculate the autocorrelation function for the Markov
chain sampling of the observable $B$ for different choices of pivot site
distribution.

As $B$ is either 0 or 1, we can write down a closed form
expression for its variance in terms of its expectation:
\begin{align}
  \mathrm{var}(B) &= \langle (B- \langle B \rangle^2)^2 \rangle =
      \langle B^2 \rangle - \langle B \rangle^2
       = \langle B \rangle - \langle B \rangle^2.
\end{align}
Then, following \cite{Li1995CriticalExponentsHyperscaling}, 
we define the autocorrelation function for 
the time series measurement of our observable $B$ as
\begin{align}
\rho_B(t) &= \frac{\langle B_s B_{s+t} \rangle - \langle B
  \rangle^2}{\mathrm{var}(B)}.
\end{align}
The integrated autocorrelation time for $B$, $\tau_\mathrm{int}(B)$,
is given in terms of
$\rho_B(t)$ as
\begin{align}
\tau_\mathrm{int}(B) &= \frac{1}{2} + \sum_{t=1}^{\infty} \rho_B(t),
\end{align}
which then enters the expression for the standard deviation of the
estimate of the expectation of $B$ after 
$n_{\text{sample}}$ Markov chain time steps:
\begin{align}
\mathrm{stdev}(\langle B \rangle) &=
    \left(\frac{2\tau_{\mathrm{int}}(B)\mathrm{var}(B)}{n_{\text{sample}}}\right)^{\frac{1}{2}}.
    \label{eq:stdev}
\end{align}
$\tau_\mathrm{int}(B)$ may be thought of as the number of Markov
chain time steps to reach an effectively new configuration with respect
to $B$.

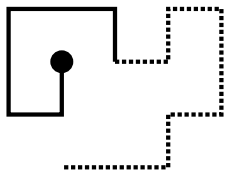
\begin{figure}[htb]
\begin{center}
\begin{tikzpicture}[ultra thick,scale=0.70]
\root
\sww \d \l \u \u \r \r \d \ew
\begin{scope}[shift={(1,0)}]
\sdw \r \u \r \d \d \l \d \l \l \ew
\end{scope}
\end{tikzpicture}
\end{center}
\caption{Minimal trapped walk of seven steps on the square lattice (solid line) with a possible
extension (dashed line). \label{fig:trapped}}
\end{figure}

It is clear from Fig.~\ref{fig:concatenation} that the shape of each of
the walks close to the joint is crucially important with respect to the
probability of intersection, whereas the shape of the walks at their far
ends will have almost negligible effect on the intersection probability.
In fact, if on the square lattice either walk is like that of
Fig.~\ref{fig:trapped}, then an intersection \emph{must} occur
regardless of the shapes of the remainders of the walks.  In
\cite{Clisby2013CalculationConnectiveConstant} we argued that sampling
pivot sites uniformly at random would lead to configurations like that
in Fig.~\ref{fig:trapped} being frozen for $O(N)$ Markov chain time
steps, and this in turn would lead to $\tau_\mathrm{int}(B) = O(N)$.

In~\cite{Clisby2013CalculationConnectiveConstant} we argued that
sampling pivot sites uniformly at all length
scales with respect to the distance to the concatenated ends would
dramatically reduce the integrated autocorrelation time, and
conjectured that in this case 
$\tint(B) = O(N^p \log N)$.
We will further test this assumption that scale-free moves drastically reduce the
integrated autocorrelation time by 
directly calculating the
autocorrelation function, and also by estimating the integrated
autocorrelation time.

We calculated the autocorrelation function for three separate 
choices of pivot site distribution. In each case we initialized the system as follows:
\begin{enumerate}
\item Use the pseudo\_dimerize procedure of
    \cite{Clisby2010Efficientimplementationpivot} to generate
two initial $N$-step SAW configurations.
\item Initialize Markov chain by performing at least $20 N$ successful
pivots on each SAW. Pivot sites are sampled uniformly at random. The
stopping criterion must be based on the number of attempted pivots so as
not to introduce bias.
\end{enumerate}
Our sampling procedure for the uniform pivot site distribution case was:
\begin{enumerate}
\item Select one of the two walks uniformly at random.
\item Select a pivot site on this walk by selecting a pivot site
    uniformly at random in the interval $[0,N-1]$.
\item Attempt pivot move, applied to the free end of the walk; update walk if result is self-avoiding.
\item Calculate $B(\omega_1,\omega_2)$, and update our estimate of $\btn$.
\item Repeat.
\end{enumerate}
The procedure with log uniform sampling was:
\begin{enumerate}
\item Select one of the two walks uniformly at random.
\item Select a pivot site on this walk by generating a pseudorandom
    number $x$ between 0 and $\log (N+1) $, and let pivot site $j =
        \lfloor e^x - 1
        \rfloor$, so that $j \in [0,N-1]$.
\item Attempt pivot move, applied to the free end of the walk; update walk if result is self-avoiding.
\item Calculate $B(\omega_1,\omega_2)$, and update our estimate of $\btn$.
\item Repeat.
\end{enumerate}
Finally, the procedure with log uniform sampling plus global rotations
(which we denote log+) was:
\begin{enumerate}
\item Select one of the two walks uniformly at random.
\item Randomly pivot each of the walks around the
    innermost sites, i.e. those with label 0. (These
    pivot moves are always successful.)
\item Select a pivot site on this walk by generating a pseudorandom
    number $x$ between 0 and $\log N $, and let pivot site $j =
        \lfloor e^x 
        \rfloor$, so that $j \in [1,N-1]$.
\item Attempt pivot move, applied to the free end of the walk; update walk if result is self-avoiding.
\item Calculate $B(\omega_1,\omega_2)$, and update our estimate of $\btn$.
\item Repeat.
\end{enumerate}

We refer to the log and log+ sampling distributions as ``scale-free''
because pivot sites are sampled uniformly at all possible length scales
with respect to the distance to the concatenation sites.

\begin{figure}[htb]
\begin{center}
\includegraphics[height=0.4\textwidth]{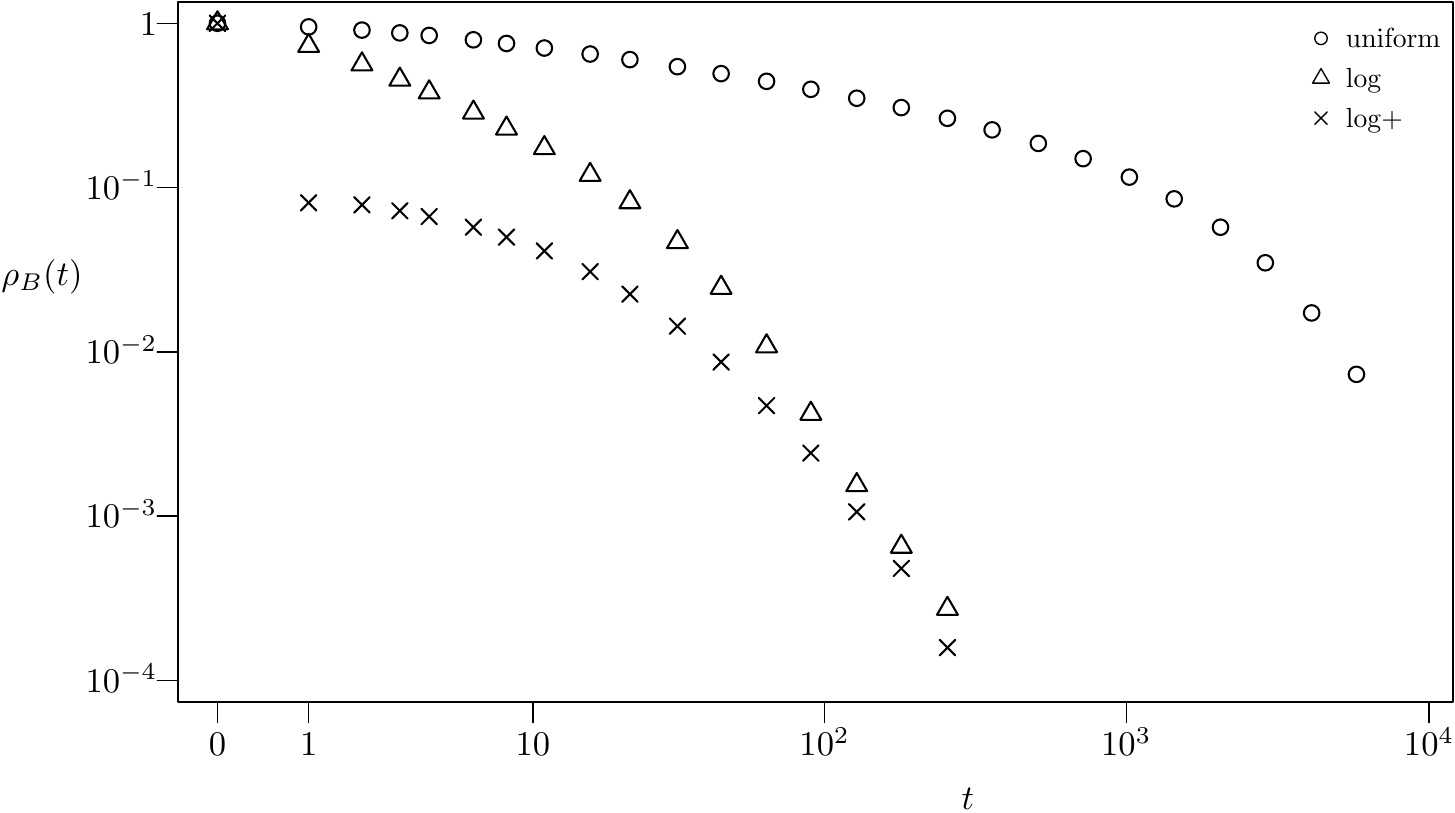}
\end{center}
\vspace{-4ex}
    \caption{Autocorrelation function $\rho_B(t)$ for 
    uniform and logarithmic
    choices of pivot site distribution for $N=999$.
\label{fig:shortautocorrelation}}
\end{figure}

\begin{figure}[htb]
\begin{center}
\includegraphics[height=0.4\textwidth]{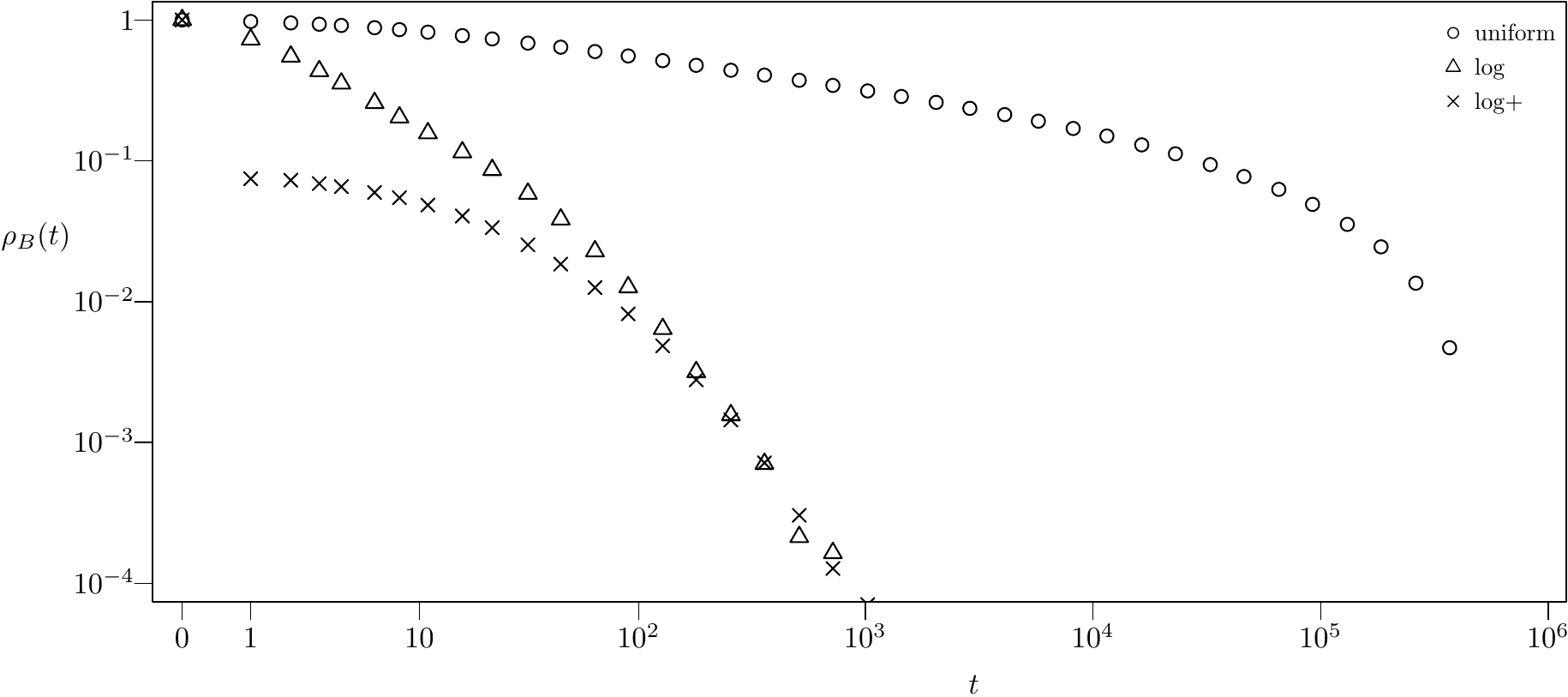}
\end{center}
\vspace{-4ex}
    \caption{Autocorrelation function $\rho_B(t)$ for 
    uniform and logarithmic
    choices of pivot site distribution for $N=99\;999$.
\label{fig:longautocorrelation}}
\end{figure}

We calculated the autocorrelation function $\rho_B(t)$ for the uniform,
log, and log+ procedures, for walks of length $N=999$ (1000 sites) and
$N=99\;999$ ($100\;000$ sites), by running simulations of the Markov
chains, and collecting information about correlations at 40 different
time intervals between 1 and 1048576.  We make log-log plots of
$\rho_B(t)$ against $t$ for $N=999$ in
Fig.~\ref{fig:shortautocorrelation} and for $N=99\;999$ in
Fig.~\ref{fig:longautocorrelation}, so that we can see the behaviour
over many time scales simultaneously. For regimes where $\rho_B(t)$ is
decaying as a power law $t^s$ with $s < 0$, we expect that the
plot will be linear with slope $s$, whereas when $\rho_B(t)$ is
decaying exponentially we expect to see the plot sharply decreasing, as
$\log \rho_B(t) = O(t) = O(\exp(\log t))$ which implies that $\log
\rho_B(t)$ will grow exponentially rapidly towards negative infinity as
a function of $\log t$.

In Figs~\ref{fig:shortautocorrelation} and \ref{fig:longautocorrelation}
it can be seen that when pivot sites are selected uniformly the
autocorrelation function decays slowly until $t$ is of the same order as
$N$ (i.e. to within a constant factor), and then decays exponentially. It
is possible that $t=O(N)$ is the only important timescale for this
Markov chain.  For the log sampling procedure, we see rapid decay which
appears to be approaching a straight line, and so is consistent with a
power law.  Decay in the autocorrelation function is dramatically faster
than for uniform sampling, as expected. Finally, for the log+ sampling
scheme we see a dramatic drop for the first Markov chain time step, due
to the use of global rotations which causes initially rapid
decorrelation, and thereafter it decays in a similar manner to the log
sampling scheme. In fact, for large $t$ we expect that $\rho_B(t)$ will
be the same for log and log+, as for long times it becomes increasingly
likely that global rotations have occurred for each walk under the log
sampling procedure. Thus log and log+ will behave similarly in terms of
asymptotic performance, but the steep initial drop in the
autocorrelation function makes it clear that log+ will better by a
not-insignificant constant factor for lengths which are accessible to
computer experiments.

\subsection{Details of computer experiment}

To extract information about $\gamma$ from (\ref{eq:bn}) we must
estimate $\btn$ in the large $N$ limit in order to reduce the influence
of corrections-to-scaling.  We sample pairs of self-avoiding walks using
the pivot algorithm and we invest computational resources approximately
uniformly on a wide range of length scales, from $N=1023$ to
$N=3355443$.  The situation is quite
different for the calculation of $\mu$
in~\cite{Clisby2013CalculationConnectiveConstant}, for which a
near-optimal design for the computer experiment required almost all
computational effort to be expended on sampling short walks.

The log+ procedure was very similar to the method used for the main
computer experiment. However, the main computer experiment was slightly
sub-optimal in two ways: (a) it was possible for the log uniform
sampling to select the sites labeled 0, and (b) one of the two global
pivot moves allowed for the identity symmetry. Each of these differences
result in slightly worse performance, and for future numerical
experiments the log+ procedure will be used (unless a procedure that is
better still can be devised).

The computer experiment was run for 200 thousand CPU hours on Dell
PowerEdge FC630 machines with Intel Xeon E5-2680 CPUs (these were run in
hyperthreaded mode which gave a modest performance boost; 400 thousand
CPU thread hours were used). In total there were $1.60 \times 10^6$
batches of $10^8$ attempted pivots, and thus there were a grand total of
$1.60 \times 10^{14}$ attempted pivots across all walk sizes.

\section{Results and analysis}
\label{sec:results}

We report our results for $\btn$ in Table~\ref{tab:data} of
Appendix~\ref{sec:data}.

In Fig.~\ref{fig:tint} we plot estimates for $\tau_{\mathrm{int}}(B)$
which we obtain indirectly from (\ref{eq:stdev}), inferring it from
batch estimates of the error in Table~\ref{tab:data}.  The accuracy of
this technique relies on the assumption that the batch error estimate is
accurate, which in turn relies upon the degree of correlation between
successive batches being negligible. For the batch sizes of $10^8$ used
in this computer experiment this condition is undoubtedly satisfied.  In
the plot of $\tau_{\mathrm{int}}(B)$ we see, remarkably, that over the
range of $N$ plotted $\tint(B)$ is growing less quickly than $O(\log
N)$! This is significantly smaller than the $O(N^p \log N)$ behaviour
postulated in our earlier
paper~\cite{Clisby2013CalculationConnectiveConstant}. It may be that
Fig.~\ref{fig:tint} does not capture the asymptotic regime, perhaps due
to the steep initial decline in $\rho_B(t)$ which is apparent for the
log+ procedure in Figs~\ref{fig:shortautocorrelation} and
\ref{fig:longautocorrelation}. However, it is possible, perhaps even
plausible, that $\tint(B) = O(\log N)$, and it certainly seems highly
probable that $\tint(B) = o(N^p \log N)$.

\begin{figure}[htb]
\begin{center}
\includegraphics[width=0.5\textwidth]{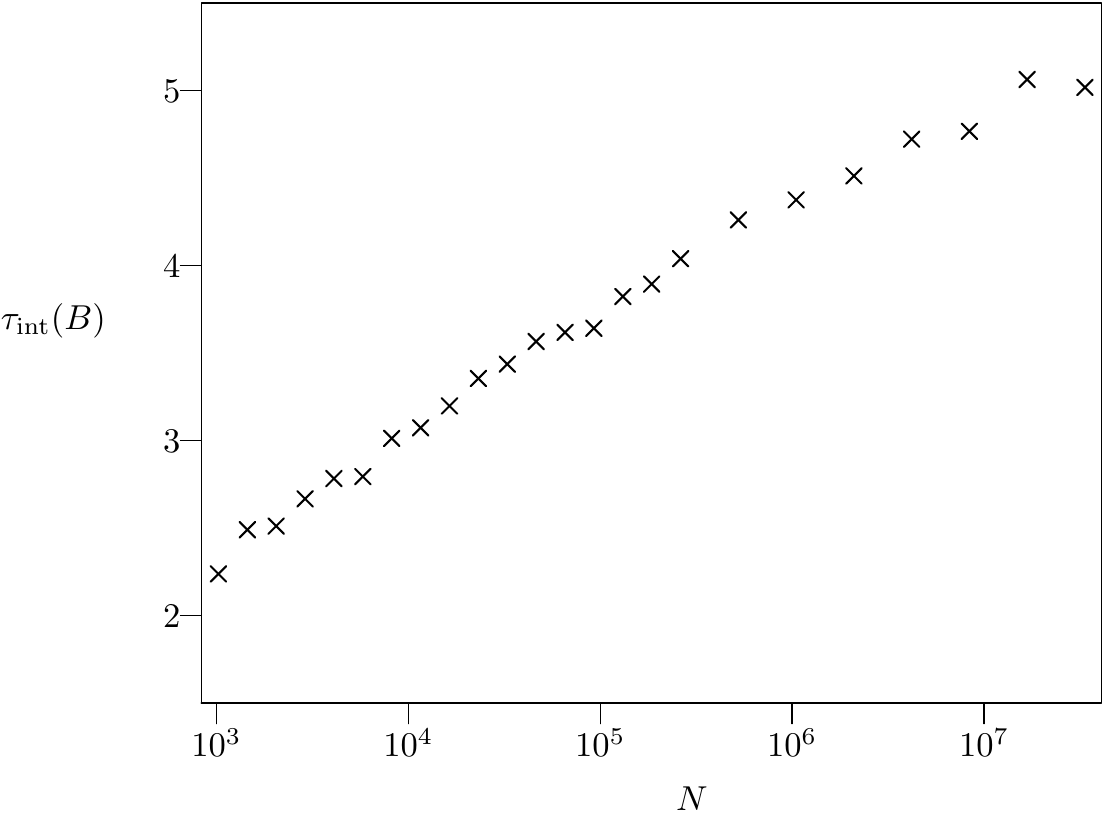}
\end{center}
\vspace{-4ex}
    \caption{Integrated autocorrelation time, $\tau_{\mathrm{int}}(B)$. These
    data are from the full Monte Carlo computer experiment and are
    calculated via the batch method.
\label{fig:tint}}
\end{figure}

We now proceed to analyze our data for $\btn$ to extract estimates for
the critical exponent $\gamma$ and amplitude $A$ via
(\ref{eq:asymptotic}). We utilize an improved observable, similarly
to~\cite{Hasenbusch20073dDilutedIsingImprovedObservable,Hasenbusch2010Finitesizescaling},
and more recently~\cite{Clisby2016HydrodynamicRadiusForSAWs}. The idea
is to combine our estimates for $\btn$ with estimates from another
observable, so as to create a new improved observable for which the
amplitude of the leading correction-to-scaling term is negligible. For
this purpose we use the estimates of the ratio of the mean-squared
end-to-end distance and the mean-squared radius of gyration, 
$\langle R_{\mathrm{E}}^2 \rangle_N / \langle R_{\mathrm{G}}^2 \rangle_N$,
from Table IV of Appendix B of~\cite{Clisby2016HydrodynamicRadiusForSAWs}.
Note that $N$ in that table refers to the number of sites, whereas here
our $N$ refers to the number of steps, which is one fewer, and so these
lengths are in one-to-one correspondence.

The expected asymptotic form of this ratio is
\begin{align}
    \frac{\langle R_{\mathrm{E}}^2 \rangle_N}{\langle R_{\mathrm{G}}^2 \rangle_N} &=
    \frac{\DDE}{\DDG} \left(1 + \frac{d}{N^{\Delta_1}} + O\left(\frac{1}{N}\right) \right).
    \label{eq:rerg}
\end{align}
Note that asymptotically this ratio of observables is a pure number,
namely the universal amplitude ratio $\DDE/\DDG$.
We now form the observable $\btn (\langle R_{\mathrm{E}}^2 \rangle_N / \langle R_{\mathrm{G}}^2
\rangle_N)^\kappa$, which involves an arbitrary
constant $\kappa$ which we will choose a value for shortly. From
(\ref{eq:bn}) and (\ref{eq:rerg}) we determine the asymptotic
form of our new observable:
\begin{align}
\btn \left(\langle R_{\mathrm{E}}^2 \rangle_N / \langle R_{\mathrm{G}}^2
\rangle_N \right)^\kappa
&= \frac{2^{\gamma-1}\mu}{A} N^{1-\gamma} \left(1 +
\frac{b}{N^{\Delta_1}} + O\left(\frac{1}{N}\right)\right)
    \left[ \frac{\DDE}{\DDG} \left(1 +
    \frac{b}{N^{\Delta_1}} + O\left(\frac{1}{N}\right) \right)
    \right]^\kappa
\\    &= \frac{2^{\gamma-1}\mu}{A}  \left(\frac{\DDE}{\DDG}\right)^\kappa N^{1-\gamma} \left(1 +
    \frac{b-d\kappa}{N^{\Delta_1}} + O\left(\frac{1}{N}\right) \right)
\\    &= K N^{1-\gamma} \left(1 +
    \frac{b-d\kappa}{N^{\Delta_1}} + O\left(\frac{1}{N}\right) \right),
    \label{eq:improved}
\end{align}
taking $K = (2^{\gamma-1}\mu / A) ( \DDE/\DDG )^\kappa$.
Thus it becomes apparent that if we choose $\kappa$ judiciously so that
$b - d \kappa \approx 0$, then our observable will have negligible
leading-order correction-to-scaling. In this case we say that the new
observable is ``improved'' with respect to the original observable
$\btn$.

\begin{figure}[htb]
\begin{center}
\begin{minipage}{0.45\textwidth}
\begin{center}
\includegraphics[width=1.0\textwidth]{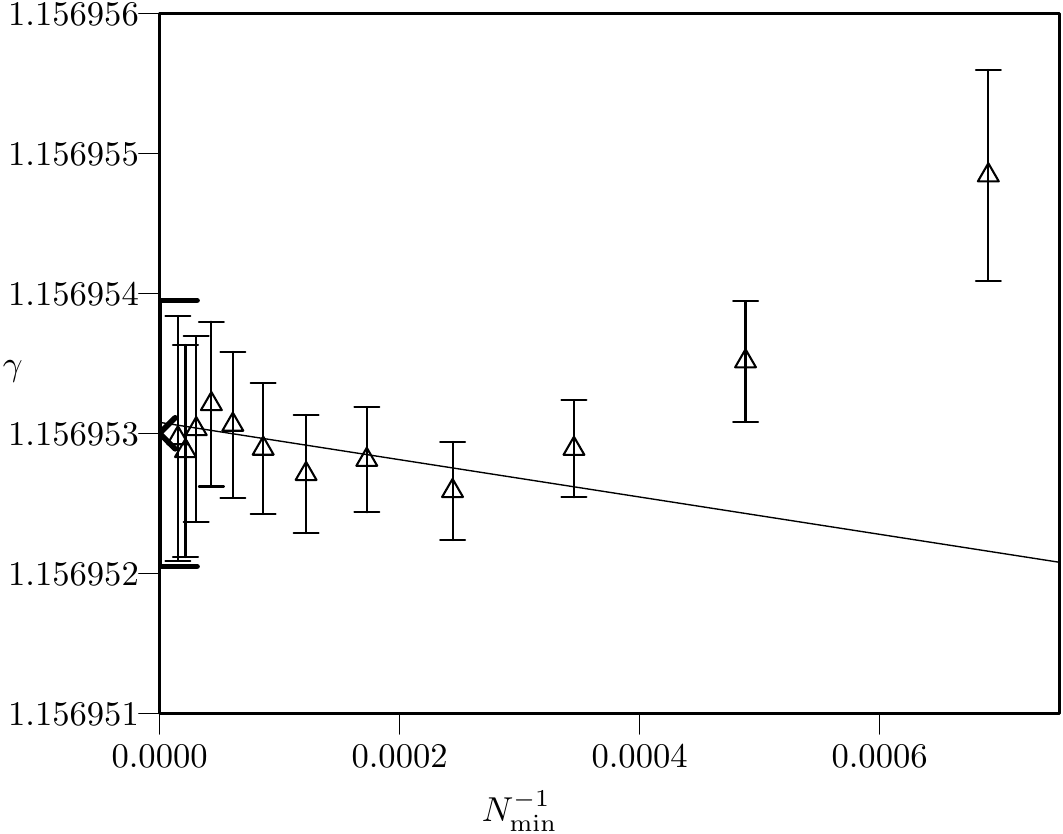}
\end{center}
\vspace{-4ex}
\caption{Estimates of the critical exponent $\gamma$, with the weighted
    least squares linear fit of the last six values shown. Our best estimate
    $\gamma=1.15695300(95)$ is shown
    in bold on the $y$-axis.
\label{fig:gamma}}
\end{minipage}
\hspace{2em}
\begin{minipage}{0.45\textwidth}
\begin{center}
\includegraphics[width=1.0\textwidth]{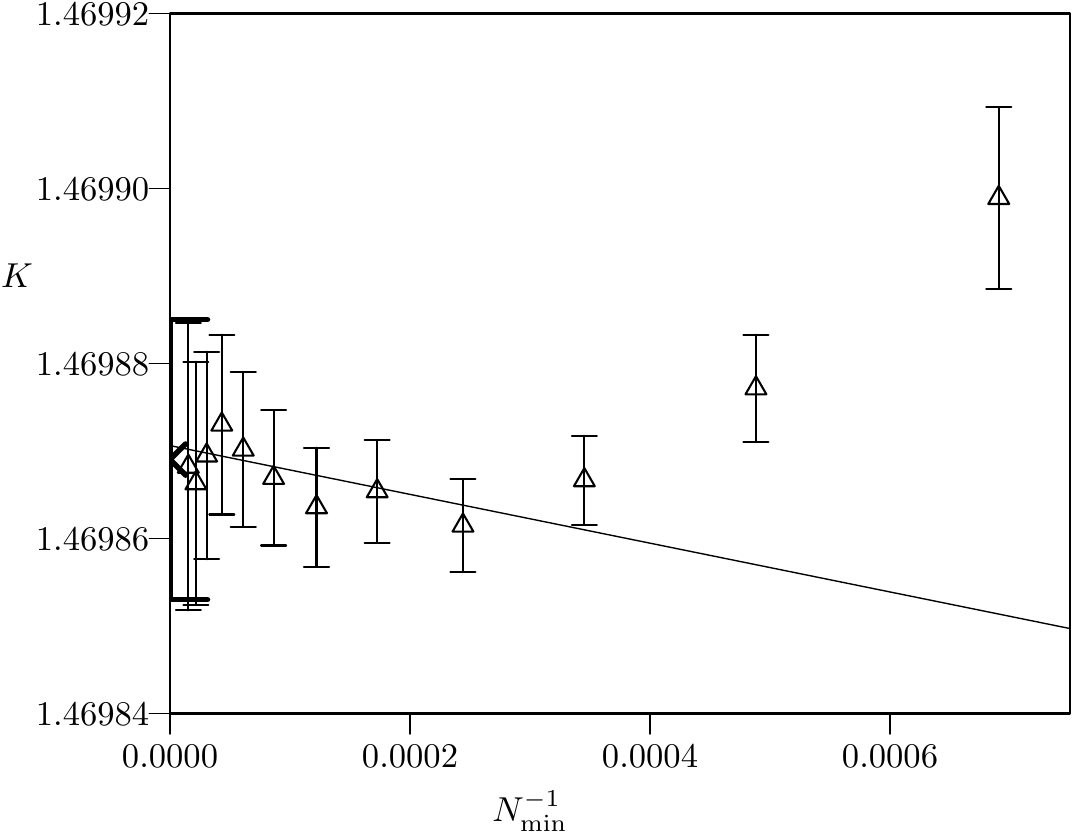}
\end{center}
\vspace{-4ex}
\caption{Estimates of the amplitude $K$, with the weighted
    least squares linear fit of the last six values shown. Our best estimate
    $K=1.469869(16)$ is shown
    in bold on the $y$-axis.
\label{fig:amplitude}}
\end{minipage}
\end{center}
\end{figure}

Our analysis was completed as follows.  We fixed $\kappa$ at an
arbitrary value (initially 0), and calculated estimates of the new
observable $\btn (\langle R_{\mathrm{E}}^2 \rangle_N / \langle
R_{\mathrm{G}}^2 \rangle_N)^\kappa$, with confidence intervals, from our
data for $\btn$ in Table~\ref{tab:data} of Appendix~\ref{sec:data}, and
the data for $\langle R_{\mathrm{E}}^2 \rangle_N / \langle
R_{\mathrm{G}}^2 \rangle_N$ in Table IV of Appendix B
of~\cite{Clisby2016HydrodynamicRadiusForSAWs}.  We then performed
weighted non-linear fits of these data using the statistical programming
language R, where our statistical model was a single power law of the
form $\text{const.} N^x$. We truncated our data by only fitting values
with $N \geq N_\mathrm{min}$, varying $N_\mathrm{min}$ to get a sequence
of estimates for which we expect the systematic error due to unfitted
corrections-to-scaling to decrease. To determine a near-optimal choice
of $\kappa$, we varied $\kappa$ and calculated the reduced $\chi^2$ for
these fits, eventually settling on a value of $\kappa = -0.585$ for
which the reduced $\chi^2$ was approximately one for all
$N_{\mathrm{min}} \geq 2895$.  These fits with $\kappa = -0.585$ gave a
sequence of estimates for $1-\gamma$ (which we converted to estimates of
$\gamma$) and $K$ from (\ref{eq:improved}). We plot these estimates in
Figs~\ref{fig:gamma} and \ref{fig:amplitude} respectively, against
$N_{\mathrm{min}}^{-1}$ as this is the expected order of magnitude of
the systematic error. This choice of variable for the $x$-axis should
result in linear convergence as the asymptotic regime is reached; we
extrapolate the fits from the right to where they intersect the $y$-axis
which corresponds to the $N_{\mathrm{min}} \rightarrow \infty$ limit.

We have extrapolated these sequences of estimates to obtain $\gamma =
1.15695300(95)$ and $K=1.469869(16)$.  Using our estimates for $K$ and
$\gamma$, together with estimates of
$\mu=4.684039931(27)$~\cite{Clisby2013CalculationConnectiveConstant} and
$\DDE/\DDG = 6.253531(10)$~\cite{Clisby2016HydrodynamicRadiusForSAWs},
we obtain $A = 1.215783(14)$. The dominant contribution to the error of
this estimate comes from $K$.

We note that analysis of results from a previous computer experiment
with poorer statistics gave $\gamma = 1.156957(9)$, where the method of
analysis used the non-improved observable $\btn$. This is consistent
with but much less precise than the final estimate reported here.  This
unpublished value was used in the estimation of critical exponents
$\gamma_1$, for self-avoiding walks tethered to a surface, and
$\gamma_b$, for bridges~\cite{Clisby2016ThreedimensionalTAWs}.

\section{Discussion and conclusion}
\label{sec:discussion}

We compare our estimates for $\gamma$ and $A$ with previous estimates in
Table~\ref{tab:parameters}, and see that the new estimates significantly
improve on the state of the art. We make the observation that estimates
for $\gamma$ have trended downwards over time, both for the series and
Monte Carlo estimates, which is perhaps symptomatic of the fact that the
systematic influence corrections-to-scaling have diminished as data for
larger $N$ has become available. The most recent series estimates have
$N = 36$, while this paper provides Monte Carlo data up to $N =
33554431$.

\begin{table}[!ht]
\caption{Comparison of parameter estimates.}
\label{tab:parameters}
\begin{center}
\begin{tabular}{llll} 
    \hline
\multicolumn{1}{c}{Source} & \multicolumn{1}{c}{$\gamma$} &
  \multicolumn{1}{c}{$A$} \\ \hline
    This work & 1.15695300(95) &  1.215783(14) \\
    Unpublished\tablefootnote{Result of an earlier computer experiment which
    used similar methodology, but with poorer statistics and no use of
    an improved observable.} & 1.156957(9) & 1.21572(18)\\
\cite{Hsu2004Polymersconfinedbetween} MC (2004)&1.1573(2) & \\
\cite{Caracciolo1998Highprecisiondetermination} MC (1998) & 1.1575(6) & \\
    \cite{Schram2011ExactEnumerationsSelfAvoidingWalks} Series $N \leq 36$
    (2011) & 1.15698(34) & 1.2150(22)\\
\cite{Clisby2007Selfavoidingwalk}\tablefootnote{Using Eqs. (74) and (75)
    with $0.516 \leq \Delta_1 \leq 0.54$.} Series $N \leq 30$ (2007) & 1.1569(6) & 1.2154(28)\\
\cite{MacDonald2000Selfavoidingwalks} Series $N \leq 26$ (2000) & 1.1585 & \\
\cite{MacDonald1992Selfavoidingwalks} Series $N \leq 23$ (1992)& 1.16193(10) & \\
\cite{Guttmann1989criticalbehaviourself} Series $N \leq 21$ (1989) & 1.161(2) & \\
\cite{Guida1998CriticalexponentsN} FT $d=3$ (1998) & 1.1596(20)& \\
\cite{Guida1998CriticalexponentsN} FT $\epsilon$ (1998) & 1.1575(60)& \\
\cite{Guida1998CriticalexponentsN} FT $\epsilon$ bc (1998) & 1.1571(30)& \\
\hline
\end{tabular}
\end{center}
\end{table}

Besides the estimates for $\gamma$ and $A$, our other main results are
the striking evidence 
in Figs~\ref{fig:shortautocorrelation} and \ref{fig:longautocorrelation}
of the efficiency gain of scale-free sampling versus
uniform sampling, and evidence from Fig.~\ref{fig:tint} which
suggests that $\tint =
O(\log N)$ for the log+ Markov chain algorithm.

The scale-free move framework described here could be applied equally as
well to other global Monte Carlo moves besides the pivot move, in
particular to cut-and-paste
moves~\cite{Causo2002CutandPermute,Arnold2007Unexpectedrelaxationdynamics}.
We expect scale-free moves to also be useful when simulating
polymers which satisfy a geometric restriction, as has already proved to be the
case for self-avoiding walks tethered to a hard
surface~\cite{Clisby2016ThreedimensionalTAWs}.
Equally, it could be useful for the sampling of
branched polymers where the distances to internal joints introduce
additional internal length scales.

One major advantage of the scale-free approach is that it is not
necessary to decide which length scale is important. Suppose, for the
sake of argument, that for a given system optimal efficiency is attained
by performing moves at one particular length scale. Since the scale-free
framework performs moves at all length scales, including the important
one, the penalty of using the scale-free algorithm is at most $\log N$
in the integrated autocorrelation time, and $\sqrt{\log N}$ in the
error.

\section*{Acknowledgements}
  Support from the Australian Research Council under
  the Future Fellowship scheme (project number FT130100972) and
    Discovery scheme (project
  number DP140101110) is gratefully acknowledged. 
\appendix

\section{Numerical data}
\label{sec:data}

\begin{table}[!ht]
\caption{Estimates of $\btn$.}
\label{tab:data}
\begin{center}
\begin{tabular}{rrrr} 
\hline
    \multicolumn{1}{c}{$N$} & \multicolumn{1}{c}{$\btn$} &
\multicolumn{1}{c}{$N$} & \multicolumn{1}{c}{$\btn$} \tstrut \bstrut \\
\hline
    1023 & 1.4507968(16) &    65535 & 0.7536518(22)\\
    1447 & 1.3734488(17) &    92671 & 0.7137264(22)\\
    2047 & 1.3002643(17) &   131071 & 0.6759013(22)\\
    2895 & 1.2310935(18) &   185343 & 0.6401084(22)\\
    4095 & 1.1656136(19) &   262143 & 0.6061940(23)\\
    5791 & 1.1037063(19) &   524287 & 0.5436837(23)\\
    8191 & 1.0450800(20) &  1048575 & 0.4876280(23)\\
   11583 & 0.9896313(20) &  2097151 & 0.4373552(23)\\
   16383 & 0.9371139(20) &  4194303 & 0.3922662(23)\\
   23167 & 0.8874326(21) &  8388607 & 0.3518267(23)\\
   32767 & 0.8403684(21) & 16777215 & 0.3155514(23)\\
   46335 & 0.7958358(22) & 33554431 & 0.2830274(22)\\
\hline
\end{tabular}
\end{center}
\end{table}

\end{document}